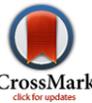

# A Model of the Effect of Uncertainty on the *C elegans* L2/L2d Decision

**Leon Avery***

Department of Physiology and Biophysics, Virginia Commonwealth University, Richmond, Virginia, United States of America

## Abstract

At the end of the first larval stage, the *C elegans* larva chooses between two developmental pathways, an L2 committed to reproductive development and an L2d, which has the option of undergoing reproductive development or entering the dauer diapause. I develop a quantitative model of this choice using mathematical tools developed for pricing financial options. The model predicts that the optimal decision must take into account not only the expected potential for reproductive growth, but also the uncertainty in that expected potential. Because the L2d has more flexibility than the L2, it is favored in unpredictable environments. I estimate that the ability to take uncertainty into account may increase reproductive value by as much as 5%, and discuss possible experimental tests for this ability.

**Data Availability:** The authors confirm that all data underlying the findings are fully available without restriction. All data are contained within the paper and Supporting Information.

**Funding:** This author has no support or funding to report.

**Competing Interests:** The author has declared that no competing interests exist.

* Email: lavery3@vcu.edu

## Introduction

The nematode *C elegans* develops from egg to adult through four larval stages, L1, L2, L3, and L4 (Figure 1). Under favorable conditions this reproductive pathway takes less than two days. However, under unfavorable conditions development follows an alternative pathway resulting in a transiently arrested third larval stage, the dauer larva. The dauer is a sort of worm spore, capable of surviving harsh conditions for a long time, and recovering if conditions improve. On recovery it becomes a superficially normal L4, with lifespan and fertility roughly the same as if it had developed through the reproductive pathway [1].

A worm must make the decision to become a dauer twice [2]. Near the time of the L1 molt, the worm decides to become either a reproductively growing L2 or a dauer-capable L2d larva. The L2 commitment to reproductive growth is irreversible at or shortly after the molt. The L2d larva, in contrast, has the option to become either a dauer or a reproductive L3. This poses a puzzle. Apparently the L2d can do anything the L2 can. Why, then, does a worm ever choose L2? Why, indeed, does the choice even exist? Yet, under favorable conditions normal worms invariably become L2s. There must be a cost to L2d development, a mechanism by which it decreases fitness under favorable conditions.

The L2d option to follow either the reproductive or the dauer pathway is valuable because the future is unpredictable. If the worm could at the L1 molt predict with perfect accuracy conditions at the end of the L2d, it could commit at the L1 molt. A worm would choose to become an L2d only if future conditions favored becoming a dauer, and the option to return to the reproductive development would never be exercised, and therefore worthless. Of course, it is not in fact possible to predict the future with perfect accuracy. Because the future is unpredictable, it is valuable to postpone the reproductive/dauer decision until a future time, when the future has become the present and is no longer uncertain.

Here I identify one possible cost of inappropriately choosing L2d. I use mathematical tools for pricing options in financial markets to estimate the value of the L2d option. This option value depends on two factors. One of these, environment quality, measures how favorable the future environment is likely to be for growth and reproduction. In fact, the L2/L2d decision is influenced by signals of food and crowding [2]. The second factor is volatility, which measures the unpredictability of the future environment. High uncertainty makes the option more valuable and therefore favors the L2d choice. Because of this dependence, an animal that can estimate uncertainty and take it into account will make better decisions than one whose decisions are based on environment quality alone. I use simple models to estimate the possible value of uncertainty information, and suggest mechanisms worms might use to acquire it.

## Results

### Reproductive value

To quantify the effects of a decision, I begin with Fisher's concept of reproductive value [3]. The reproductive value of a worm is proportional to the expected number of its descendants at some distant future time, based on the information available to the worm. The descendants of animals with high reproductive value will, by definition, be a larger part of the future population than those of animals with low reproductive value. (Note that *C elegans* are usually found as hermaphrodites and that self-fertilization is the dominant method of reproduction in the wild [4,5]. Thus there is no overlap between the descendants of two worms. In this paper I neglect the effect of the rare males that occur.) Animals that make decisions that maximize their reproductive value will therefore be favored by evolution.

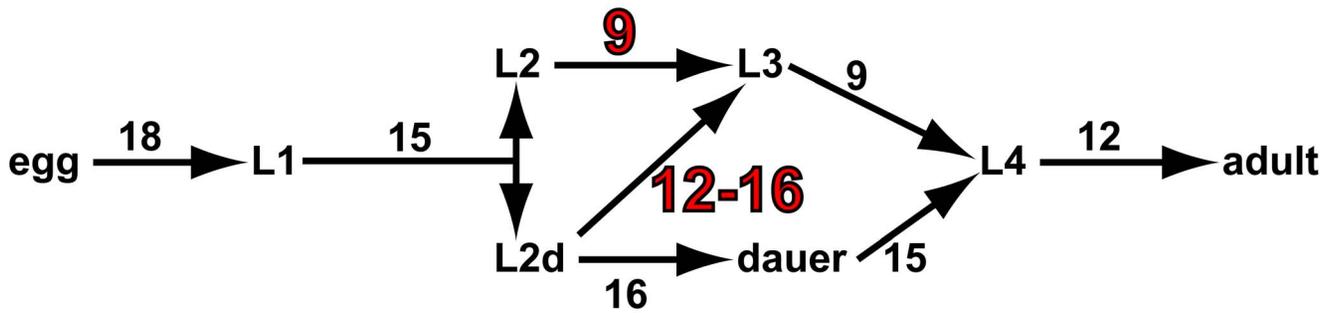

**Figure 1.** *C elegans* **developmental pathways.** This figure shows schematically the pathways a *C elegans* egg may follow to adulthood. Numbers show the approximate duration in hours of each stage in the laboratory at 20°C. The L1 stage lasts 15 h, and at the end (the L1 molt) the worm decides to become either an L2, committed to reproductive development, or an L2d, which has the option of becoming a dauer larva. This decision is the subject of this paper. The duration of the L2 and L2d pathways to the L3 are highlighted to show the 3–7 delay incurred by following the L2d pathway. A worm may remain a dauer for many months; times shown are for the development of the L2d from L1 molt to L2d molt, and for the recovery of the mature dauer.
doi:10.1371/journal.pone.0100580.g001

The value of a worm depends on its age and condition. A gravid adult hermaphrodite with 20 eggs in her uterus about to be laid has a value at least 20 times the value of a single egg in the same environment. Yet that adult was herself a single egg a few days ago. Her value must therefore have increased in that time. It increased because the value of the egg derives from its potential to become a gravid adult. The older the egg and the worm that hatches from it become, the closer it gets to adulthood, the greater its chances of escaping dangers and finding resources so as to reach adulthood. Similarly, value depends on condition. A young larva near death from starvation is less likely to reach adulthood than a larva of the same age with abundant stored nutrient reserves, so the first is less valuable than the second.

Value also depends on the environment. A young larva near death of starvation in an environment devoid of food is unlikely to have any descendants, and therefore has low value. Its value is not zero because there is a small chance that it may find food before it dies. The same starved larva is more valuable in the presence of food because it is more likely to reach adulthood, and, if the food supply is very large, because many of its children and grandchildren will reach adulthood.

Less obviously, value depends on information available to the worm through its senses or internal state. The value of the starving larva is very low, but if the larva's chemical senses inform it of food in the vicinity, its chance of survival and therefore its value rise abruptly. It might be argued that the true value (whatever that might mean) of the worm is not changed by changing information—only the worm's estimate of its value is different. Readers more comfortable with this view may want to mentally replace "value" with "expected value" or "estimated value". In any case, it is this information-based value that optimal decisions must maximize. Because value depends on information, it can change quickly.

## Dependence of value on age

Historically, most *C elegans* eggs have not reached adulthood. The argument is essentially that of Malthus. It is based on two assertions: first, that the average *C elegans* adult produces many more than one egg, and second, that just one of the children of the average *C elegans* adult reaches adulthood. The first assertion is based on *C elegans* reproductive physiology. Under ideal laboratory conditions a *C elegans* hermaphrodite produces about 300 progeny [6,7]. The mean brood size in the wild is likely to be less, but several arguments suggest that it is considerably more than 1. The capacity to produce such a large brood is achieved at a high cost: I estimate that the gonad and uterus of an adult hermaphrodite are about ¼ of her volume. It is unlikely that such a large gonad would increase fitness unless the worm actually used the reproductive capacity it affords. Furthermore, even when completely deprived of food an adult hermaphrodite can produce at least 8 progeny by "facultative vivipary", i.e., consuming her own biomass to produce but not lay eggs, which then hatch internally and eat the mother [8]. It seems safe to assume that the average number of progeny produced by an adult hermaphrodite is at least 8, and probably larger.

That, on the average, only one of these progeny reaches adulthood is clear. *C elegans* has been in existence for at least 1 million years. The mass of a *C elegans* adult is roughly 3 μg. If a population started with a single adult doubled 111 times, its mass would exceed the mass of the Earth. Thus, the mean rate of growth of the Earth's *C elegans* population has been less than 1 doubling per 9,000 years. By similar reasoning, the population cannot have dwindled faster than one halving per 9,000 years. From this it can be deduced that the mean number of descendants of an adult *C elegans* hermaphrodite that reached adulthood in six months (the maximum plausible generation time) has been between 0.999962 and 1.000038. Even if one assumes a very recent huge expansion of the *C elegans* population these constraints are only slightly relaxed. For instance, a 100-fold expansion of the *C elegans* population in the last 100 years would require at most that the average adult give rise to 1.023 adults. In this paper I assume that populations are precisely at steady-state. This assumption is not necessary—the models described below also work for non-steady-state populations and lead to similar conclusions—but the exposition is simplified. The steady-state assumption together with the evidence that an adult produces more than eight eggs implies that at least seven in eight *C elegans* eggs fail to become fertile adults in the wild.

In a steady-state population, the value of an egg or a larva is proportional to the probability that it will become an adult. In one simple model, the probability of failing to advance in age, i.e., of dying or permanently arresting development, is a constant per unit time, which I call λ, the discount rate. (In reality, of course, λ may vary—the assumption that it is constant is a modeling simplification.) I have estimated λ by several different methods (see λ: discount rate in Methods), which give values from 0.027 h$^{-1}$ to $0.068\ \mathrm{h}^{-1}$. The method I consider most reliable yields $\lambda = 0.042\mathrm{h}^{-1}$. This means that a worm of developmental age $a$ hours has probability 0.042 of failing to reach age $a+1$ and probability 0.958 of advancing one hour. More generally, the probability that a worm of age $a$ reaches age $A$ is $P(a,A) = e^{-\lambda(A-a)}$. The value of an age $a$ worm is proportional to $e^{\lambda a}$. This simple model makes the most sense for a worm committed to a single developmental pathway. An L2 larva, for instance, has no path to adulthood except by developing through every hour of L2.

These arguments suggest a solution to the puzzle of the L2d, and a method of quantifying the cost. Golden and Riddle [2] found that it takes 3–7 hours longer for a worm to develop from L1 to L3 via the L2d than through the L2 (Figure 1). One plausible reason for this is that the dauer stores fat in order to survive for months without food [9], and the L2d must therefore take the time to eat more. Thus an L2d must run the gauntlet of a dangerous world for up to 7 hours longer than one that takes the L2 pathway. A mutation that eliminated the L2d pathway, forcing the worm to always follow the L2d pathway, would suffer a reduction in value by a factor of between $e^{-3\lambda}$ and $e^{-7\lambda}$ in a good environment (0.88–0.74, using the estimate above). Such a worm would be at a serious disadvantage to wild-type in good environments and would have no advantage in poor environments.

## Binary model

The value of a worm with options, e.g. an L2d which can choose to follow either the reproductive pathway or the dauer pathway, cannot be so simply characterized by a single probability of developmental advance. Its value is derived from the values of the choices it is free to make. A simple example illustrates how this works. Imagine a world in which there are only two possible environments, good or bad (Table 1). In this world, an L2 about to molt into an L3 has no value in a bad environment (it always dies without progeny). The value of a dauer depends less on the environment than that of other stages. For simplicity, the model assumes it is entirely independent of the environment. The world can vary in two ways: the value of the new L3 relative to the dauer and the probability of the good environment may change.

Table 1A considers the most predictable variation, in which the future is certain: a good environment with probability 1. Since the value of the dauer is independent of environment, it is convenient to express the value of the L2 about to molt in terms of dauers. In this example I suppose it is worth 1 dauer in the good environment. An L2d about to molt will become a dauer when things are bad and an L3 when things are good, so its value is 1 dauer in both the bad and good environments. The worm must choose between L2 and L2d at the time of the L1 molt. If it chooses the L2 pathway, its chance of making it to the L2 molt, which takes 9 h, is $e^{-9\lambda} \approx 0.68$. Its value is 1 dauer if it reaches the L2 molt, 0 if it doesn't, for a discounted mean of 0.68. If the L1 chooses the L2d pathway its chance of making it to the next molt is $e^{-16\lambda} \approx 0.51$ (assuming the maximum possible delay of 7 hours, for a total of 16). There it will be worth 1 dauer whichever choice it makes, for a discounted mean of 0.51. Thus, because of the L2d delay, the L2 is the better choice.

However, value depends on information. Suppose the worm doesn't yet know whether the environment will be good or bad, but only that they occur with equal probability 0.5 (Table 1B). Suppose also that in this world, the good environment is better than in example A, so that the value of the new L3 is 2 dauers. In this world, the question "What is the value of the L2 just before the molt?" has three different answers, depending on what informa-tion the worm has. A worm that knows that the environment is good has value 2, and a worm that knows the environment is bad has value 0. The value of a hypothetical worm that hadn't yet found out which environment it was in, however, would be the mean of these weighted by their probabilities: $0.5 \times 0 + 0.5 \times 2 = 1$. Similarly, an L2d will choose dauer (value 1) in the bad environment and L3 (value 2) in the good. Its value, if it hadn't yet found out what kind of environment it was in, would be $0.5 \times 1 + 0.5 \times 2 = 1.5$ dauers. In this toy example, when the environment is uncertain, the option to choose between dauer and L3 is worth half a dauer.

In fact, by the time of the molt an L2 or an L2d will know whether the environment is good, but it can't decide between L2 and L2d then. The real relevance of the informationless values is when the decision is made, at the time of the L1 molt. As before, these values must be discounted by the probability of successfully reaching the next molt. Thus, the L2d choice is worth $0.51 \times 1.5 = 0.76$ dauers at the L1 molt, the L2 $0.68 \times 1 = 0.68$, and L2d is preferred.

Table 1C shows a third variation of this example, in which the good environment occurs with 0.667 probability and the L2 is worth 1.5 dauers at the L2 molt. At the L1 molt both the L2d and the L2 are worth 0.68 dauers, so the worm is indifferent.

These examples were contrived to make a point. In all three cases the value of the L2 choice is the same: 0.68. Also, in all three cases, the value of the dauer is the same: 1 at the L2d molt, or 0.51 at the L1 molt. Yet the optimum choice differs among the three. This shows that to make the best decision, it is not enough to know whether, on the average, the animal will have more future descendants on the reproductive pathway than the dauer pathway. Another piece of information is needed. This piece of information is the uncertainty of the predicted future. High uncertainty favors the worm that has options over a committed worm. Thus in case B, with high uncertainty, the L2d is favored over the L2, while in case A, with no uncertainty, the L2 is favored over the L2d. If the worm knows only the average quality of the future environment (as measured by the relative values of the L3 and dauer), it can't always make the optimum choice. But if it also knows the uncertainty (as measured, for instance, by the standard deviation of the value of the L3), it can always choose optimally.

An example may help to clarify the meanings of value and uncertainty. In nature, *C elegans* is thought to alternate between brief periods of rapid population growth, which occur when an animal finds a food source such as a dead snail or a rotting apple, and long droughts, which the worm survives as a dauer [10]. An individual that is lucky enough to find a rotting apple early can hope to found an exponentially expanding population—this is facilitated by the ability of a self-fertilizing hermaphrodite to reproduce alone and the explosive growth rate of an unrestrained *C elegans* population. We don't know how big such a local clonal population can grow, but considering the relative sizes of a worm and an apple, it is not implausible that a rare lucky worm gives rise to thousands or even a million dauers before the food is exhausted. Since, as argued above, the world-wide *C elegans* population is close to steady-state, it is inevitable that most of these dauers die without reproducing. In fact, if the mean number of dauers produced by a dauer that finds a food patch is $N$, the probability that one of these dauers similarly finds a food patch is $1/N$.

Thus, consider two worms. One is an L1 in a rotting apple that has been nearly exhausted. It is teeming with worms and other invertebrates, and the food will be gone in 2–3 days. This worm's best strategy is to grow to adulthood and produce progeny, and it almost certainly can do that before the food runs out. The second worm is an L1 that hatched after the food had run out and

**Table 1.** Binomial model for L2, L2d, and dauer value

**A.**

| State | Value at L2/L2d molt if environment is... | | | Value at L1 molt if environment is... | | |
|---|---|---|---|---|---|---|
| | Bad ($p = 0.0$) | Good ($p = 1.0$) | Mean Value | Bad ($p = 0.0$) | Good ($p = 1.0$) | Mean Value |
| dauer | 1 | 1 | 1 | 0.51 | 0.51 | 0.51 |
| L2 | 0 | 1 | 1 | 0 | 0.68 | 0.68 |
| L2d | 1 | 1 | 1 | 0.51 | 0.51 | 0.51 |

**B.**

| State | Value at L2/L2d molt if environment is... | | | Value at L1 molt if environment is... | | |
|---|---|---|---|---|---|---|
| | Bad ($p = 0.5$) | Good ($p = 0.5$) | Mean Value | Bad ($p = 0.5$) | Good ($p = 0.5$) | Mean Value |
| dauer | 1 | 1 | 1 | 0.51 | 0.51 | 0.51 |
| L2 | 0 | 2 | 1 | 0 | 1.37 | 0.68 |
| L2d | 1 | 2 | 1.5 | 0.51 | 1.02 | 0.76 |

**C.**

| State | Value at L2/L2d molt if environment is... | | | Value at L1 molt if environment is... | | |
|---|---|---|---|---|---|---|
| | Bad ($p = 0.333$) | Good ($p = 0.667$) | Mean Value | Bad ($p = 0.333$) | Good ($p = 0.667$) | Mean Value |
| dauer | 1 | 1 | 1 | 0.51 | 0.51 | 0.51 |
| L2 | 0 | 1.5 | 1 | 0 | 1.03 | 0.68 |
| L2d | 1 | 1.5 | 1.33 | 0.51 | 0.76 | 0.68 |

doi:10.1371/journal.pone.0100580.t001

proceeded to the L1 starvation diapause. This worm has only a $1/1{,}000$ probability of finding an apple before it dies, but if it finds one, it can expect to produce 1,000 fertile adults. Both of these worms have the same value: that of 1 adult. But their uncertainty is very different. Consequently, the L2d option is more valuable to the second worm than to the first.

The binary model is unrealistically simple, but some of its properties are quite general: an option is more valuable in an uncertain environment than a predictable one, and therefore, to make the optimal choice between the L2 and the L2d, the L1 should take into account not only how good or bad the future is likely to be, but its uncertainty about that future.

## Continuous models

The binomial model allows for only two decision times, L1 molt and L2/L2d molt, and two environments, good and bad. For more realistic models whose quantitative predictions might be a useful approximation to reality, I looked for inspiration in finance. A dauer is like cash. Its value is relatively stable and independent of the environment. A reproductive larva is analogous to a share of stock. Its value represents the best estimate of future growth prospects. Like a worm, a share of stock can grow exponentially in value in a favorable economic environment, coming eventually to be worth hundreds or thousands of times its original value, or the company can die out and the stock become worthless. The value of stock depends on information and can thus change quickly. Some stocks are relatively stable in value, while others are more volatile. An L2d is like a call option on a stock: it allows but does not obligate the future exchange of a fixed cash price (one dauer) for a share of stock (an L3).

Black and Scholes [11] modeled stock value as a geometric Brownian motion: a random variable whose logarithm is normally distributed, with a variance that increases linearly with time. This model oversimplifies reality and fails in some important cases, but it is a useful approximation much of the time, and is widely used in pricing financial options. In the Black-Scholes model, like the simpler binary model, the value of an option depends on two things, the value of the assets whose exchange it enables (the dauer and the L3, in the case of an L2d), and volatility, which determines how uncertain the current estimate of those future values is.

The Black-Scholes model can be applied to the L2d (see Fixed-time European model in Methods), but it fails to describe the biological problem in two important ways. First, financial options have specific expiration dates. In contrast, the time at which an L2d can choose between reproductive and dauer pathways depends on developmental age, and the rate at which a worm develops is variable, depending on such things as temperature and food supply. The effect of this difference is to increase unpredictability, since the L1 not only doesn't know how the environment will change—it doesn't even know how long it has to make its decision. Second, the Black-Scholes model assumes that an option cannot be used before its expiration. However, an L2d need not wait until the molt to make its decision—it can commit to the reproductive pathway before the molt [2,12]. The ability to decide early is valuable because it decreases the developmental delay associated with the L2d decision to as little as 2–3 hours, and therefore its cost.

Figure 2 shows how value depends on the quality of the environment in a model that takes these factors into account. The L2d value curve is plotted for four different values of uncertainty: 0 (a completely predictable environment), 0.5, 2, and ∞ (a maximally unpredictable environment). The line representing the value of an L2 is also shown, using a measure of environment quality such that L2 value does not depend on uncertainty. The L2 line crosses the zero uncertainty curve at an environment quality of 0.3 dauers, i.e., where the expected value of an L2 is equivalent to 0.3 dauers. Thus, in a completely predictable environment, the L1 should choose the L2 pathway if the environment is good enough that the L2 will produce progeny worth, on the average, 0.3 dauers. If the environment is worse than that, the L1 should instead choose the L2d pathway. In more uncertain environments the L2d option has more value, and the crossing occurs at higher quality values. Thus, at a modest uncertainty of 0.5, meaning that environment quality will typically change by 0.67–1.5-fold between the L1 molt and the L2d molt, the L1 should require an L2 value of 0.45 dauers to persuade it to take the L2 pathway, and at an uncertainty of 2, the threshold is 0.97 dauers. In a maximally uncertain environment, the L2 must have a value of 2.4 dauers (beyond the right-hand-edge of the graph) for the L2 to be optimal. The threshold environment quality thus varies as much as eight-fold, depending on the level of uncertainty. The phase diagram Figure 3 summarizes the optimal strategy based on both environment quality and uncertainty.

## Low and high uncertainty extremes

The L2d value curve has a particularly simple form at low uncertainty and at high uncertainty. In a low uncertainty environment the future is completely predictable. The worm already knows at the L1 molt what decision it will make at the L2 molt. Indeed, it knows whether or not it will exercise the option to switch to the L2 pathway during L2d development. There are only two possible future courses. In one, the worm follows the L2d pathway all the way to the molt, then becomes a dauer. Its value, if it follows this pathway, is simply $e^{-16\lambda} V_{\text{dauer}}$, the value of a new dauer discounted by the cost of 16 h development. In the other, the worm follows the L2d pathway only until the first possible time at which it can switch to the L2 pathway. Following this pathway, it takes 3 h longer than the 9 that would have been required to reach L3 than if it had chosen the L2 pathway from the start, so its value is $e^{-(9+3)\lambda} V_{L3}q$. ($V_{L3}q$ is the value of an L3 at environment quality $q$.) A switch from L2d to L2 later than the earliest possible time would cause more delay, and therefore lower value, so will never be optimal. Since the L2d will choose the most valuable of these two futures, its value is just the maximum of the two. That is, it is flat at $e^{-16\lambda} V_{\text{dauer}}$ up to the point at which it intersects $e^{-(9+3)\lambda} V_{L3}q$, and then it follows the latter's linear increase.

The high uncertainty curve is more surprising and requires more explanation. In an extremely uncertain environment, the value of an L2d is the *sum* of the values of its two options, the L3 and the dauer, discounted by the developmental delay. How can one worm have the value of two? The answer is shown by the high-uncertainty example given above: an L1 with a $1/1{,}000$ chance of value 1,000. If the $1/1{,}000$ chance pays off, this worm follows the reproductive pathway, thus capturing all the value of the L2 option. If the environment goes bad, it chooses to become a dauer. Since this occurs 99.9% of the time, the L2d also captures 99.9% of the value of the dauer option. Thus, the L2d's value is the value of the L2 option plus almost all the value of the dauer option.

L2d value at intermediate uncertainty in Figure 2 and the strategy curve in Figure 3 depend on detailed assumptions of the model, which specify the nature of environmental variation and the timing of development in the wild. Strategies based on the binary model or the Black-Scholes model have similar features, but the exact shape is different. Surprisingly, however, the curves for high and low uncertainty are independent of these model assumptions. It is easy to understand why detailed assumptions would not matter in the low uncertainty environment: in this world, the future is completely predictable. But the high uncertainty limit is also independent of detailed assumptions. The reason, it turns out, is that there is basically only one way of having very high uncertainty. The value of a worm can never be negative, so it can never decrease by more than 100%. Since the value can never decrease by more than 100%, all the high volatility has to occur on the upside. But since by definition the average value of the future possibilities must equal the current value of the worm, high uncertainty can only mean that high future values occur with very low probability, and low future values occur with probability close to 1. In other words, in a high-uncertainty environment, almost all the value of a population lies in very rare worms that achieve extraordinary reproductive success. The L2d has nearly the value of both its L2 and its dauer options in any such environment.

## Costs of ignoring uncertainty

The previous sections showed that to make the optimal L2/L2d decision, the L1 must take into account both environment quality and uncertainty. But how much difference does it make? How much less fit is a worm that ignores uncertainty? It is impossible to answer this question without knowing what kind of variability worms experience in the wild. However, the high and low uncertainty limits allow calculation of an upper bound.

Using uncertainty in decisions is valuable only if uncertainty varies. The world in which uncertainty matters the most is one in which very high uncertainty and very low uncertainty both occur, and each occurs with high probability. Assume, therefore, that a worm finds itself either in a low uncertainty or a high uncertainty environment with equal probability. In this world, compare the value of two types of worms: a Smart worm that bases its decisions on both the quality and the uncertainty of its environment, and a Dumb worm that bases its decisions solely on quality. The optimal strategies for such worms are shown in Table 2. The optimal strategy when ignoring uncertainty is to use an environment threshold intermediate between the low uncertainty and the high uncertainty thresholds.

Figure 4A compares the value curves for the Smart and Dumb worms. Even though differences in uncertainty can result in an eight-fold difference in the L2/L2d threshold, the cost of ignoring uncertainty is comparatively small. Figure 4B plots the cost in value, and Figure 4C as a percent of the value. In very poor environments and very good environments, where the optimal decision is independent of uncertainty, there is no cost to ignoring it. The cost is largest in mediocre environments where the L2/L2d decision is more difficult, reaching a maximum of over 5% of value.

In conclusion, taking uncertainty into account when making the L2/L2d decision may increase value by up to 5%.

# Discussion

## Evaluation of the model

The main conclusion of this paper is that a worm that takes uncertainty into account will make better decisions than a worm that ignores it. These better decisions may increase the worm's contributions to future generations by as much as 5%. This conclusion is based on a model that, like all biological models, grossly oversimplifies reality. Some defects of the model are real, and some only apparent.

Among the latter is the apparent assumption that the value of a reproductively developing worm depends only on environment quality, and not on uncertainty. Surely a more unpredictable environment is worse, all else equal? In fact, the model is consistent with this intuitive insight. The apparent independence of L2 value and uncertainty is a consequence of choosing a measure of quality that, by definition, *includes* uncertainty. This can be seen most clearly in the Binomial model example (Table 1). The "predicted environment quality" $q$ is identical in all three cases, as shown by the fact that the mean value for the L2 is the same. But it is the same only because greater uncertainty (in C, for instance) is compensated for by a higher upside value.

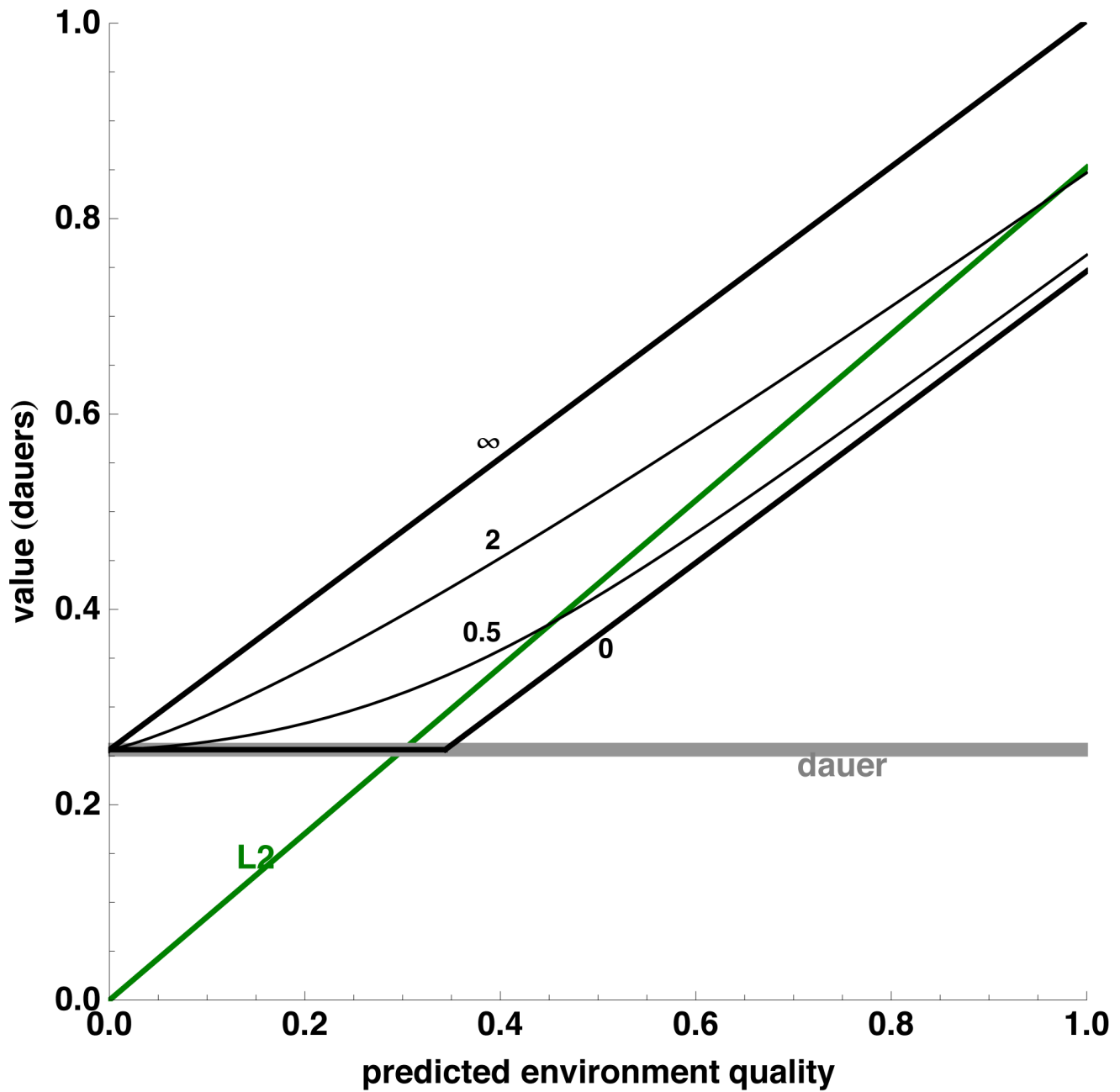

**Figure 2. Volatility and option value.** Plotted in black is the value immediately after the L1 molt of an L2d that can choose between L3 and dauer, calculated with the Hybrid model described in Methods. Green and gray curves show the values of committed L2 and dauers at the same time, respectively. (The committed dauer is hypothetical—normal worms do not commit to dauer at the L1 molt.) Environment quality is measured by the value of an L3 larva. Black lines plot value for environments of different volatility, quantified as described in Methods by uncertainty, a number related to the factor by which environment quality will typically change during L2d development. The two heavy black lines show value when the future environment is completely predictable (uncertainty = 0) or completely unpredictable (uncertainty = ∞). Thinner black lines plot value against quality for intermediate levels of uncertainty 0.5 and 2.
doi:10.1371/journal.pone.0100580.g002

An example might help to clarify this. Consider a single person buying a house. She cares about both the price, and the size. However, in the end she can only purchase one house, and her best strategy is to select that house in which she expects to be happiest. If one defines her expected level of future happiness, based on the price and size of the house, as the "predicted house quality", then she can make her decision solely on the basis of this variable. There is nothing interesting or insightful about this claim—it is a tautology, based on the way "predicted house quality" is defined.

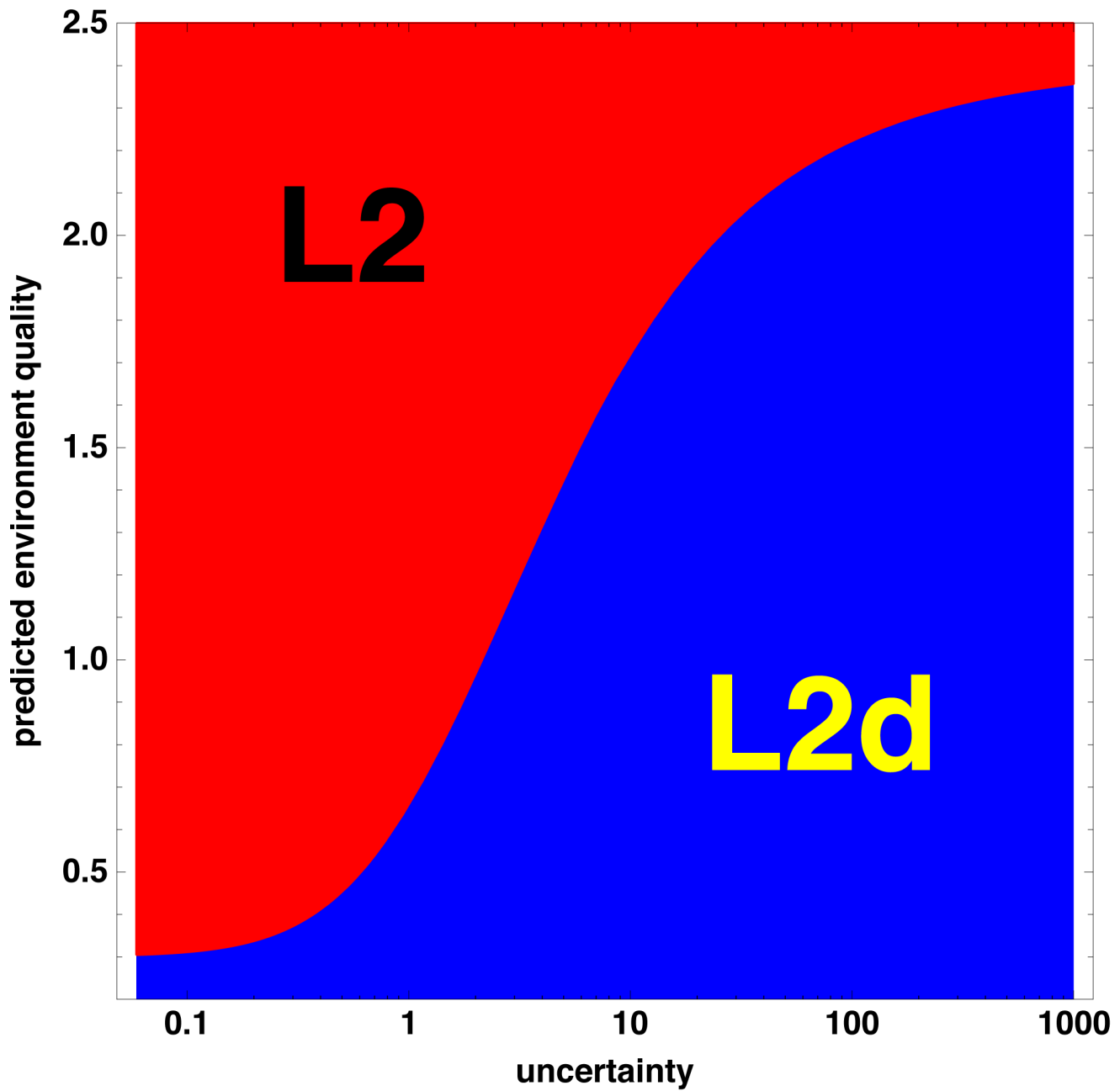

**Figure 3. L2/L2d decision curve.** This phase diagram shows the optimal decision strategy for the L1 choosing between L2 and L2d pathways on the basis of uncertainty and predicted environment quality, as measured by the value of the L2. L2d is favored in high uncertainty and poor environments (the blue region), L2 in low uncertainty or good environments (red).
doi:10.1371/journal.pone.0100580.g003

It becomes nontrivial if we also consider another shopper, a man with a family. He also cares about price and size. Shopper 1's "predicted house quality" might be valuable to him in making his decision. But his choice cannot be based solely on the "predicted house quality" defined for her. This family man will care more about size. If he knew *both* the quality *and* the size of every house, he could make a better decision. For instance, he might choose the highest quality house that has at least three bedrooms.

The claim that an L1 needs to know both environment quality and uncertainty is something like this, but stronger. The L2 and the L2d are like the two shoppers. Even if we define a number, "predicted environment quality" that combines everything that determines the L2's future, this is not enough to figure out how good the future will be on the L2d pathway. Another piece of information is needed, uncertainty. But there is a closer relationship between the L2 and the L2d than between the two shoppers, because the L2d has the option of becoming the future L2. In the example one might argue that Shopper 1's evaluation is really of little relevance to Shopper 2. But there is no question that the value of an L2 is relevant to the value of an L2d.

Another apparent oversimplification is the attribution of an explicit option to the L2d, but not to the L2, the L3, or the dauer. Obviously all animals have options. And these options improve their chance of survival and therefore increase their value in

**Table 2.** Smart and Dumb worm strategies.

| A. | Dumb worm strategy | |
|---|---|---|
| environment | $q < 1.2$ | $q > 1.2$ |
| choice | L2d | L2 |
| | | |
| B. | Smart worm strategy | |
| | low uncertainty | |
| environment | $q < 0.3$ | $q > 0.3$ |
| choice | L2d | L2 |
| | high uncertainty | |
| environment | $q < 2.4$ | $q > 2.4$ |
| choice | L2d | L2 |

doi:10.1371/journal.pone.0100580.t002

uncertain environments. The dauer, for instance, can at any time recover, or remain a dauer. Even the L2 can slow down growth and development if food is scarce, and has the possibility of becoming an adult that may lay eggs, entering adult reproductive diapause, or undergoing matricide [13]. The L2d, however, clearly has more options than the L2, because it can become an L2, but it can also do something else. It is for this reason that the L2d is predicted to become more valuable relative to the L2 in a more uncertain environment. In the simplified models used here, the ability of the dauer to recover means that the dauer value curve is not a flat line as shown in Figure 2, but curves up at high values. This has no effect on the L2/L2d decision, since the dauer will only be chosen in a poor environment. The options available along the L2 pathway are not explicitly modeled, but are assumed to contribute to the environment quality, which is defined on the basis of the value of a reproductively developing worm. It would of course be interesting to develop more complicated and realistic models in which some or all these options in the *C elegans* life cycle are explicitly modeled.

This modeling choice has a consequence for experimental tests. It is not enough to determine whether uncertainty affects the L2/L2d decision. In fact, it is likely that uncertainty affects all developmental and diapause decisions, since all explicitly or implicitly affect future options. The strong prediction of the model is more subtle: The L2/L2d decision should be more sensitive to uncertainty than the L3/dauer decision or the roughly inverse dauer recovery decision. That is, the strategy curve for the L3/dauer decision should lie below and to the right of the curve for the L2/L2d decision (Figure 3).

## Hedging

A true defect of all the models considered here is that they assume an individual worm has a reproductive value independent of other worms in the population. (Equivalently, the value of the population is assumed to be the sum of the values of the worms in it.) In such a model, bet hedging never makes sense: every worm should do whatever maximizes its mean contribution to future generations based on the information available to it, and all worms with the same information should behave identically. This assumption is justified if the random variations experienced by the worms in a population are uncorrelated with each other ("demographic stochasticity" [14]). If, however, every worm in the population experiences identical random variation ("environmental stochasticity" [14]), it may be optimal for the genotype if some worms make individually non-optimal decisions. The clearest case is decisions that risk death. A decision that risks the death of an individual may be good for that individual, if it also has substantial upside possibilities. However, in a population subject to pure environmental stochasticity, a decision that risks the extinction of the population is never optimal, no matter how high the upside may be [14]. In reality, of course, stochasticity is neither purely demographic nor purely environmental. The randomness experienced by different members of a population is correlated, but not identical. Probably, uncertainty is more environmental for short times, before a population has time to disperse, and more demographic for longer times.

It is likely that the worm does in fact hedge dauer decisions. The most striking evidence for this is the effect of ascaroside pheromones on dauer formation. In standard assays it has proven impossible, even at very high pheromone concentrations, to force 100% of worms to go dauer [2,15,16]. Even at concentrations 1000 times those required to induce >10% dauer formation, some worms do not become dauers [15]. (Interestingly, since it *is* possible to get 100% reproductive development, this suggests counterintuitively that in nature dauer development is the dangerous choice.)

The models used here assume pure demographic stochasticity, and therefore cannot account for hedging. It is intuitively clear, however, that the main conclusion of this paper, that uncertainty is important in optimal decision-making, holds even if there is environmental stochasticity. Hedging is a response to uncertainty, albeit only uncertainty of a particular type. If there is no uncertainty, there should be no hedging, since the value of the population is reduced by some animals choosing predictably suboptimal strategies. By assuming pure demographic stochasticity, I make the weakest possible assumption. Yet, even in this case, accounting for uncertainty allows better decisions.

## How uncertain is the environment?

The inference that a worm will make better decisions if it takes uncertainty into account depends on the assumptions that (1) the environment experienced by a lineage of worms is uncertain, and (2) uncertainty varies a lot from one environment to another. The first assumption is not controversial. For instance, Felix and Braendle [10] refer to worms' "boom-and-bust lifestyle exploiting ephemeral resources", and state, "Critical life-history choices likely reflect adaptations to the fluctuating and ephemeral natural habitat of *C. elegans*." Considering the size of food concentrations in the wild relative to the worm, uncertainties of the level at the right edge of Figure 3, i.e. rare worms experiencing environments in which their lineage can expand 1,000-fold, do not seem unrealistic.

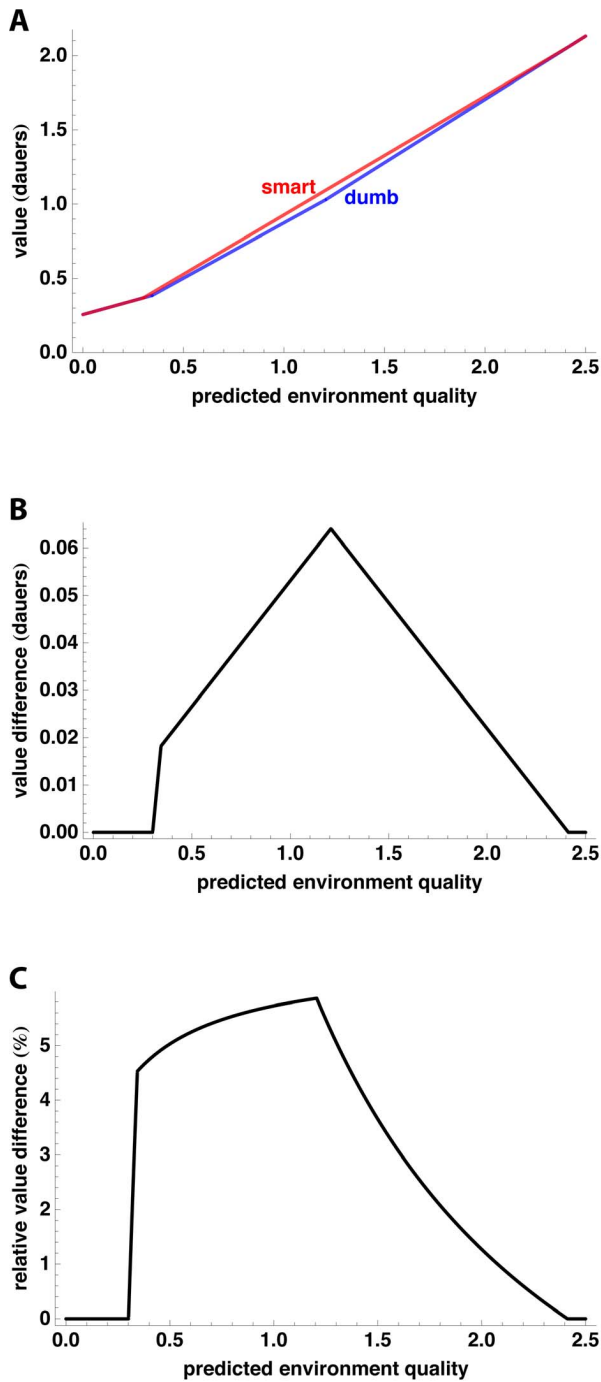

**Figure 4. Cost of ignoring uncertainty. A**. The value of a worm following either the Smart or the Dumb worm strategy of Table 2. **B**. The difference in value between Smart and Dumb worm strategies. **C**. Difference in value as a percent of the Smart strategy.
doi:10.1371/journal.pone.0100580.g004

However, there is almost no information on how much uncertainty varies. Certainly, there must be *some* variation. For instance, viability, fertility, and growth rate are all affected by temperature, and *C elegans*, a cosmopolitan species, is commonly found in temperate climates where temperatures fluctuate seasonally above and below the optimum. Similarly, rotting organic matter is more reliably found at some seasons and locations than others. However, as shown in Figure 3, the effects of uncertainty will only make a big difference if there are also relatively predictable environments, with uncertainties on the order of 1 or less (i.e., where future value typically varies by less than a factor of 2). Furthermore, uncertainty is unimportant in very good or very poor environments (Figure 3, Figure 4). We have no information on the frequency or even existence of conditions of medium environment quality and varying uncertainty, and the prospect of measuring them in field studies is daunting.

I estimated that the use of uncertainty information might increase the value of a worm by as much as 5%. This is an upper bound, and the typical value benefit will be smaller. Reproductive value is not the same as fitness, but they are related: relative fitness is proportional to the per-generation difference in value. If, for instance, a lineage of worms finds itself in a situation where the use of uncertainty information confers a 5% advantage once every ten generations, the fitness benefit is 0.5%. Compared to known selective effects acting on survival traits, this is a respectable number and might well drive evolution [17–19]. (For example, Hoekstra et al. [18] find that reported values of $\beta_\sigma$, the relative fitness difference corresponding to one standard deviation of a quantitative trait, are distributed exponentially with a median of 8.8%. From this it can be estimated that 33% of reported effects have $\beta_\sigma < 5\%$ and about 4% have $\beta_\sigma < 0.5\%$. $\beta_\sigma$ includes the combined selective effect of variation caused by the environment and all loci controlling a trait. Recent results suggest that quantitative traits are often controlled by many genes with small individual effects [20]. This suggests that, at the level of an individual gene, even very small fitness differences can be important.) If, on the other hand, such conditions occur only once in a thousand generations, the fitness benefit is only 0.005%, a tiny effect (though still capable in theory of being important in a large population over long times). It is also true that there are many other options in *C elegans* development—other decisions whose accuracy may be improved by accounting for environmental uncertainty. Thus the value of sensory and computational engines capable of estimating uncertainty might well go beyond the specific circumstances in which they aid the L2/L2d decision.

### Do worms use uncertainty?

These considerations suggest that worms might estimate uncertainty and use these estimates in the L2/L2d decision and others. How might they do this? Clearly, the worms can't directly measure future uncertainty. If they are to estimate uncertainty, they must do so using a proxy, such as past uncertainty or spatial variation, that is correlated with future uncertainty. There are at least four possibilities.

The first is the trivial answer: the estimate may be genetically fixed, and optimized by evolution. This would correspond to drawing a fixed vertical line on Figure 3, and making the L2/L2d decision based only on estimated environment quality. This is the "Dumb worm" strategy of Figure 4. The remaining three strategies all involve some active method of evaluating uncertainty.

The second method was hinted at in Results. The worm may estimate uncertainty based on its correlation with environmental characteristics. For instance, a worm in a rotting apple in which the food is about to be exhausted can foresee its future with considerable precision. A starving worm searching for food in a target-rich environment may have the same value, but it is much less able to predict it. The second worm should therefore favor L2d more than the first.

The third method is the most obvious: a worm can estimate its uncertainty of the future based on the volatility of its past. That is, an animal that has grown up in a varying environment might favor L2d, while a worm that grew up in a constant environment would favor L2. These decisions might depend on the worm's own experience, or (epigenetically) on its mother's.

The fourth method is communication with others. Worms communicate using ascaroside pheromones, which affect the L2/L2d and L3/dauer decisions, as well as other behaviors [2,21]. We know now that the hypothesis of Golden and Riddle [2] that dauer pheromone is a constitutively secreted population density signal, while correct, is an oversimplification. *C elegans* and other nematodes release several different ascarosides, the ascarosides released depend on the worm's condition and environment, and they respond differently to different pheromones [21–23]. This suggests that pheromones may serve not just to detect crowding, but to pool information about the environment. Suppose, for instance, that a worm that experiences a good environment releases one pheromone, while a worm that experiences a bad environment releases a different pheromone (e.g. C9 ascarosides released by well-fed L1 cultures, compared to C5 released by starved cultures [22]). An L1 would interpret good pheromone as evidence of a good environment and the bad as evidence of a bad environment. A mixture of good and bad would suggest a variable and uncertain environment, and bias the L2/L2d decision towards L2d. It is also possible to combine the third and fourth mechanisms: a worm that experiences a variable environment might release a pheromone that directly signals volatility.

The first of these hypotheses, that the L2/L2d decision is based on a genetically fixed estimate of uncertainty, is essentially untestable, since it is negative. The second, that uncertainty is estimated based on its correlation with environmental characteristics, does make an experimental prediction: that the L2/L2d decision will depend differently on environment than do the L3/dauer and dauer recovery decisions. Unfortunately, this prediction is not specific to the uncertainty hypothesis; one can easily propose alternative explanations.

Hypotheses three and four, however, can be experimentally tested. Hypothesis three implies that a worm that has experienced a volatile past (or whose mother has experienced a volatile past) will be more likely to choose L2d than a worm that has experienced a constant past. This prediction can be tested by manipulating the L1's experience of temperature, food, and pheromone (factors known to influence the dauer decision) between hatching and the L1 molt. For instance, one group of worms might be given a constant, low density of food from hatching, while a second group received, in alternation, no food for time $t_1$, followed by high density food for time $t_2$, with $t_1$ and $t_2$ chosen so that, at the L3/dauer decision, the fraction of worms that choose dauer is the same in the two groups. Hypothesis three then predicts that the second group of worms will choose L2d more often than the first group. By choosing different food densities and different alternation regimes, the phase diagram of Figure 3 could be explored.

The clearest evidence for hypothesis four would be identification of the proposed good and bad pheromones. There is however an experimental test short of this, a mixing experiment. It begins by isolating conditioned medium from worms in good conditions (well-fed, optimal temperature, uncrowded) and from worms in bad conditions (starved, high temperature, crowded). The first prediction is that these conditioned media would influence the L2/L2d and L3/dauer decisions: good CM would favor reproductive development, bad L2d or dauer. The second, more important prediction is that a mixture of good and bad CMs would bias the

L2/L2d decision more strongly towards L2d than it does the L3/dauer decision towards dauer.

These two tests are not trivial, but seem feasible with current technology.

## Methods

### Reproductive Value

To determine the best way to make the dauer decision, I follow other authors (e.g. Houston and McNamara [14]) in assuming that every worm has a reproductive value, and that the optimal decision is the one that maximizes this value. The concept of reproductive value (henceforth simply "value") was introduced by Fisher [3], who, considering the effects of age on selection in human populations, defined it as the answer to the question, "To what extent will persons of this age, on the average, contribute to the ancestry of future generations? The question is one of some interest, since the direct action of Natural Selection must be proportional to this contribution." Since *C elegans* reproduces primarily by self-fertilization, so that each animal has only one parent, a worm's value may be defined more simply as a number proportional to the average number of its descendants at some distant future time. The focus on the distant future overcomes a weakness of simply counting the worm's children: their contribution to future generations depends not just on their number, but also on their age, condition, and access to resources. If, however, one waits long enough for descendants to disperse and the environment to revert to its mean, these transient factors should average out.

The words "on the average" in Fisher's definition are important. The number of future descendants of a single worm does not approach a finite limit: it is well-known that if you wait long enough in a uniparental population, the lineages of all but one animal will die out, and the descendants of one particular animal will take over the entire population [24]. It is only by averaging over animals that are similar in some way that one can hope to define value. Since Fisher was concerned with age, he averaged over all persons of a given age and sex, and the value he calculated was therefore a function of age and sex.

Here I am concerned with developmental or behavioral decisions. Any decision must be based on the information available to the worm at the time of the decision. Therefore value is based on the average expected number of future descendants of an animal, with the average taken over all animals in the population that have the same information.

$$v(\mathcal{I}) = \lim_{T \to \infty} \mathbb{E}[C(T)|\mathcal{I}] \qquad (1)$$

$C(T)$ (for clan) is the number of descendants of a worm at time $T$. $\mathcal{I}$ is a σ-algebra representing the information available to the worm. It includes information about the worm's internal state, e.g. its age and condition, as well as information about the present and past environment gathered by the senses. $\mathbb{E}[C(T)|\mathcal{I}]$, a conditional expectation, is thus the expected or average number of descendants at time $T$ of a worm with the information $\mathcal{I}$. Definition (1) assumes that the population eventually reaches steady-state, so that the limit as $T \to \infty$ is defined. Non-steady-state populations present no essential difficulty, but a more complicated definition is necessary. For simplicity, this paper deals only with populations at steady-state.

Value so defined is a function of information. Since information can change rapidly, as rapidly as the senses operate, value can change rapidly. In this way the value of an animal is like the value of a financial security such as a share of stock, which is also an estimate of future growth, depends on information, and can change suddenly in response to news.

The information available to a worm changes with time, and is therefore represented by a filtration $\mathcal{F}(t)$. Substituting $\mathcal{F}(t)$ for $\mathcal{I}$ in (1) gives

$$v(t) = v(\mathcal{F}(t)) = \lim_{T \to \infty} \mathbb{E}[C(T)|\mathcal{F}(t)] \qquad (2)$$

$v(t)$ so defined is a martingale—the best estimate of its future average is its current value.

## Binomial model

The analogy between reproductive value and the value of a financial security suggests that models used to value securities might also be applied to worms. In particular, an L2d, which may become either an L3 or a dauer larva, is like a financial option, which confers the right but not the obligation to purchase a stock (the L3) at a predetermined strike price (the value of one dauer) at some future time (the L2d molt). (To be more precise, the L2d is like the combination of a call option on a stock plus cash equivalent to the strike price.) The binomial model [25], a simple illustrative model often used to demonstrate the properties of options, can also be applied to the L2d. Although it is too simple to be quantitatively accurate, it is easy to understand and reproduces the qualitative features of more realistic models. In this model there are only two possible future environments: good and bad. The good environment ensues with probability $p$, the bad with probability $1-p$. The value of an L3 is $v$ in the good environment and 0 in the bad. The value of the dauer is 1 in both. The L2d, which can choose to become either a dauer or an L3, has value $\max(1,v)$. If there is some cost to becoming an L2d, its value is instead $d\max(1,v)$, where $0 < d \le 1$ is a discount factor.

## Fixed-time European model

Black and Scholes [11] developed a model for valuing financial options that is realistic enough to make useful quantitative predictions. In this model stock price fluctuations are modeled as a geometric Brownian motion, and the option can only be exercised at a fixed time (a European-style option). An analogous model can be developed for the L2d by measuring environment quality $q(t)$ by the ratio of the value of a reproductive L3 larva to the value of a dauer. This makes sense, since being relatively insensitive to environment is the whole point of the dauer larva. Following Black and Scholes [11], $q(t)$ is modeled as a geometric Brownian motion with volatility $\sigma$, governed by the stochastic differential equation (SDE)

$$dq(t) = \sigma q(t) dW_1(t) \qquad (3)$$

$dW_1(t)$ is a Brownian motion. (The Black-Scholes SDE usually includes an additional $rq(t)dt$ term. $r$ is the rate of population growth, zero in a population at steady-state.)

This model leads to the following formula for L2d value, similar in form to the Black-Scholes formula:

$$v(t,q(t)) = \Phi\left(\frac{\sigma^2(T-t) - 2\log q(t)}{2\sigma\sqrt{T-t}}\right) + q(t)\Phi\left(\frac{\sigma^2(T-t) + 2\log q(t)}{2\sigma\sqrt{T-t}}\right) \qquad (4)$$

where

$v(t,q(t))$ is the value of the L2d at time $t$ in units in

which a dauer has value 1

T is the time of the L2d molt $\qquad (5)$

$\Phi(z)$ is the standard normal cumulative

distribution function

As in the binomial model, the cost of becoming an L2d may be modeled by multiplying $v(s,t)$ by a discount factor $d$.

## Variable growth rate models

**Growth model.** Financial options have fixed expiration dates. The development of a worm, in contrast, depends on conditions. Development is delayed if the rate of food intake is low [26], and if food is inadequate, development may arrest, or the worm may die. To model this, I let $a(t)$ represent the worm's developmental age as a function of time. The dynamics of the state of the worm are modeled as a Brownian motion with negative drift

$$da(t) = -\alpha dt + v dW_2(t)$$
$$\alpha, v > 0 \qquad (6)$$

$\alpha$ and $v$ are parameters that determine the mean rate of growth and how much it varies. $W_2(t)$ is a Brownian motion independent of $W_1(t)$.

While not literally realistic, this growth model has several realistic properties:

- For most worms, things get worse with time.
- There is a broad range of times taken to reach adulthood (or any other specified goal).
- Some worms never reach adulthood.
- The further away a goal is, the longer it takes to reach it.
- The probability of reaching a goal decreases the further away it is. Specifically,

$$P(a,A) = e^{-\lambda(A-a)} \qquad (7)$$

is the probability that a worm of age $a$ reaches age $A > a$. $\lambda$, the discount rate, is given by

$$\lambda = \frac{2\alpha}{v^2} \qquad (8)$$

The hardest part of this model to swallow is that it appears to describe backward aging: that a worm can become developmentally younger with time. This seeming absurdity is resolved by thinking of $a(t)$ as some combination of the worm's age and condition related to its potential to reach reproductive adult stage and produce progeny. (In essence, (7) is assumed to hold *by definition*.) A decrease of $a(t)$ with time would correspond to depletion of stored nutrients, loss or damage of cellular proteins, etc.

The L2d molt is delayed compared to the L1 molt. The delay is greatest if the worm is maintained under dauer-inducing conditions for the entire interval, smallest if, immediately after the L1 molt, it is switched to conditions that favor reproductive growth [2]. This suggests that the L2d develops more slowly than the L2. I therefore model L2d growth like L2 growth, except slowed down by a delay factor $\delta$

$$da(t) = -\delta \alpha dt + \delta v dW_2(t) \tag{9}$$

**Reproductive larva.** The value of a larva committed to reproductive growth (but not a larva on the dauer pathway) is modeled as the product of two functions, one that depends only on the state of the worm, and one that depends on environment quality. Environment quality $q(t)$ represents the available information about the present and future environment on the expected number of progeny. I assume that the relevant state of the worm is summed up in the single function of time, $a(t)$, which is a combination of developmental age, nutrient reserves, and condition as described above. Thus,

$$v(t) = f(a(t))q(t) \tag{10}$$

In the financial analogy, $q(t)$ is the price of a stock and $f(a(t))$ is the number of shares held. But this analogy is no longer precise, since $f(a(t))$ fluctuates with time in a way that is not under the worm's control. Thus, the Black-Scholes formula (4) no longer holds.

$f(a(t))$ and $q(t)$ are assumed independent of each other. Since $v(t)$ is a martingale, each of $f(a(t))$ and $q(t)$ is a martingale. Itô differentiation of $f(a(t))$ and substitution of the SDE governing $a$ leads to the SDE

$$df(a(t)) = \left(-\alpha f'(a(t)) + \frac{v^2}{2}f''(a(t))\right)dt + vf'(a(t))dW_2(t) \tag{11}$$

Since $f(a(t))$ is a martingale, the $dt$ term must vanish, leading to the ordinary differential equation (ODE)

$$0 = -\alpha f'(a) + \frac{v^2}{2}f''(t) \tag{12}$$

with boundary condition

$$f(-\infty) = 0 \tag{13}$$

This has solution

$$f(a) = Ce^{\frac{2\alpha}{v^2}a} = Ce^{\lambda a} \tag{14}$$

$C$ is an arbitrary constant that determines the units of value. I define it so that a mature dauer larva has value 1. The discount rate $\lambda = \frac{2\alpha}{v^2}$ is the same as in eq (7), governing the probability of developmental progress and $f(a)$ is directly proportional to this probability. This shows that the dependence of value of $a(t)$ is entirely accounted for by the probability of advancing to later stages.

**L2d.** L2d value doesn't have the simple product form assumed for reproductive larvae. Its value is a function of $\boldsymbol{a}$ and $\boldsymbol{q}$, but is derived from its capacity to eventually become a dauer or an L3. Differentiation of $v(a(t),q(t))$, substitution of SDEs (3) and (9), and setting the coefficient of $\boldsymbol{dt}$ to 0 leads to the partial differential equation (PDE)

$$
\begin{aligned}
0 &= \sigma^2 q^2 v_{qq} - 2\alpha \delta v_a + v^2 \delta^2 v_{aa} \\
&= \sigma^2 q^2 v_{qq} - 2\alpha \delta v_a + \frac{2\alpha}{\lambda}\delta^2 v_{aa}
\end{aligned}
\tag{15}
$$

I estimate the value of the L2d in the models below by solving this PDE.

## European model

**Boundary and terminal conditions.** In the European model (named after European-style financial options), the L2d makes a single decision if and when it reaches the L2d molt to become either a dauer or an L3. In fact, this is not true—the L2d can exercise the option to commit to L3 before the molt. However, the European model is easier to solve, and its solution is the basis for the solution of the American and hybrid models below, in which early exercise is allowed.

Define $a$ so that the value of $a$ is 0 at the L2d molt, negative at earlier ages. We have boundary conditions

$$
\begin{aligned}
v(-\infty, q) &= 0 \\
\lim_{q \to 0} v(a,q) &= V_d e^{\frac{\lambda}{\delta}a} \\
\lim_{q \to \infty} \frac{v(a,q)}{q} &= V_{L3} e^{\frac{\lambda}{\delta}a}
\end{aligned}
\tag{16}
$$

These conditions hold for $a < 0, 0 \leq q$. $V_d$ is the value at the L2d molt of a worm that has committed to become a dauer, and $V_{L3}q$ is the value at the L2d molt of a worm that has committed to become an L3. The second boundary condition says that in a very bad environment $\langle q \approx 0 \rangle$, the worm will always choose to become a dauer, so it can be priced by discounting the dauer value. Likewise the last boundary condition says that in a very good environment the worm will always become an L3. In addition the terminal condition,

$$v(0,q) = \max(V_d, V_{L3}q) \tag{17}$$

holds for $q > 0$. Terminal condition (17) says that at the molt the L2d will commit to either dauer or L3 development, whichever has the highest value. This terminal condition is the primary way in which biology enters the solution.

**Transformation of the PDE.** I have not been able to find a closed-form solution for this model, but I have found an efficient numerical solution based on Fourier transforms. Begin by transforming the PDE (15). First, make the substitutions

$$
\begin{aligned}
q &= \frac{V_d}{V_{L3}}e^u \\
v(a,q) &= e^{\frac{\lambda}{2\delta}a}\sqrt{V_d V_{L3}}q\, y_1(a,u)
\end{aligned}
\tag{18}
$$

to get PDE

$$0 = -\left(\frac{\alpha\lambda}{2} + \frac{\sigma^2}{4}\right)y_1 + \sigma^2 y_{1uu} + \frac{2\alpha\delta^2}{\lambda}y_{1aa} \qquad (19)$$

with boundary conditions

$$y_1(-\infty, u) = 0$$

$$\lim_{u \to -\infty} e^{-\frac{\lambda}{2\delta}a} e^{u/2} y_1(a, u) = 1 \qquad (20)$$

$$\lim_{u \to \infty} e^{-\frac{\lambda}{2\delta}a} e^{-u/2} y_1(a, u) = 1$$

and terminal condition

$$y_1(0, u) = \max\left(e^{-u/2}, e^{u/2}\right) \qquad (21)$$

Let

$$y_{\max}(a, u) = e^{\frac{\lambda}{2\delta}a}\left(e^{-u/2} + e^{u/2}\right) \qquad (22)$$

$y_{\max}$ solves the PDE (19) and boundary conditions (20) but not the terminal condition (21). Thus,

$$y(a, u) = y_{\max}(a, u) - y_1(a, u) \qquad (23)$$

must satisfy homogeneous PDE

$$0 = -\left(\frac{\alpha\lambda}{2} + \frac{\sigma^2}{4}\right)y + \sigma^2 y_{uu} + \frac{2\alpha\delta^2}{\lambda}y_{aa} \qquad (24)$$

homogeneous boundary conditions

$$y(-\infty, u) = y(a, -\infty) = y(a, \infty) = 0 \qquad (25)$$

and terminal condition

$$y(0, u) = y_0(u) = e^{-u/2} + e^{u/2} - \max\left(e^{-u/2}, e^{u/2}\right)$$
$$= e^{-|u|/2} \qquad (26)$$

To check the final simplification, confirm that when $u \geq 0$ both expressions reduce to $e^{-u/2}$, and when $u < 0$ to $e^{u/2}$.

**Separation of the PDE.** This system can be solved by separation of variables in the usual way. Look for solutions of the form

$$y_\omega(a, u) = e^{\kappa a} e^{i\omega u} \qquad (27)$$

Substituting into the PDE (24) gives

$$0 = \alpha\left(\frac{2\delta^2\kappa^2}{\lambda} - \frac{\lambda}{2}\right) - \frac{1}{4}\sigma^2\left(1 + 4\omega^2\right)$$
$$\kappa = \pm\sqrt{\frac{\lambda(2\alpha\lambda + \sigma^2 + 4\sigma^2\omega^2)}{8\alpha\delta^2}} \qquad (28)$$

$\kappa < 0$ is inconsistent with the boundary condition $y(-\infty, u) = 0$, so only the positive root $\kappa = \sqrt{\frac{\lambda(2\alpha\lambda + \sigma^2 + 4\sigma^2\omega^2)}{8\alpha\delta^2}}$ is of interest. $y(a, u)$ must be a sum of solutions of the form

$$y_\omega(a, u) = \exp\left(a\sqrt{\frac{\lambda(2\alpha\lambda + \sigma^2 + 4\sigma^2\omega^2)}{8\alpha\delta^2}}\right)e^{i\omega u} \qquad (29)$$

Let $\tilde{y}(a, \omega)$ be the Fourier transform of $y(a, u)$,

$$y(a, u) = \frac{1}{\sqrt{2\pi}}\int_{-\infty}^{\infty}\tilde{y}(a, \omega)e^{-i\omega u}d\omega \qquad (30)$$

Letting $a = 0$ in (30), the terminal condition becomes

$$y_0(u) = \frac{1}{\sqrt{2\pi}}\int_{-\infty}^{\infty}\tilde{y}_0(\omega)e^{-i\omega u}d\omega \qquad (31)$$

$\tilde{y}_0(\omega)$ is the Fourier transform of $y_0(u)$,

$$\tilde{y}_0(\omega) = \sqrt{\frac{8}{\pi}}\frac{1}{1 + 4\omega^2} \qquad (32)$$

whence

$$\tilde{y}(a, \omega) = \sqrt{\frac{8}{\pi}}\frac{1}{1 + 4\omega^2}\exp\left(a\sqrt{\frac{\lambda(2\alpha\lambda + \sigma^2 + 4\sigma^2\omega^2)}{8\alpha\delta^2}}\right)$$
$$y(a, u) = \frac{2}{\pi}\int_{-\infty}^{\infty}\frac{1}{1 + 4\omega^2}\exp\left(a\sqrt{\frac{\lambda(2\alpha\lambda + \sigma^2 + 4\sigma^2\omega^2)}{8\alpha\delta^2}}\right)e^{-i\omega u}d\omega \qquad (33)$$

The solution $y(a, u)$ is thus the convolution of $y_0(u)$ with a kernel $K(a, u)$ whose Fourier transform is $\exp\left(a\sqrt{\frac{\lambda(2\alpha\lambda + \sigma^2 + 4\sigma^2\omega^2)}{8\alpha\delta^2}}\right)$. A discrete approximation to (33) can be computed numerically using the fast Fourier transform. Also, the existence of a solution that blows up in the negative $a$ direction makes direct numerical solution of the PDE difficult—the Fourier transform solution evades this problem.

Once $y(a, u)$ is known, $v(a, q)$ is calculated using (18), (22), and (23).

$$v(a, q) = e^{\frac{\lambda}{2\delta}a}(V_d + V_{L3}q) - e^{\frac{\lambda}{2\delta}a}\sqrt{V_d V_{L3}}q y\left(a, \log\left(\frac{V_d}{V_{L3}}q\right)\right) \qquad (34)$$

I have developed other methods for numerical solution of the European model, but the Fourier transform method is most efficient.

## American and Hybrid European/American models

**American model.** In the European model, an L2d retains the option to develop as a dauer until the L2d molt, and thus must incur the maximum delay of 7.6 h that results from growth at the slow L2d rate for that entire time. In fact, this is not correct—under some conditions an L2d may abandon the option to become a dauer before the molt, effectively switching from an L2d to an L2 [2,12]. This early switch reduces the developmental delay incurred by following the L2d pathway and is therefore advantageous in a good environment.

In the American model (named after American-style financial options), an L2d may switch to the L2 pathway at any time. If it does so, its subsequent development proceeds at the normal reproductive rate (eq (6)) rather than the slower L2d rate (eq (9)). Calculation of the value of an L2d in the American model proceeds backward from the molt in small time steps. Since at the molt there is no longer any possibility of an early switch to the L2 pathway, terminal value is identical for the American and European L2d and is given by (17):

$$v_A(0,q) = v_E(0,q) = \max(V_d, V_{L3}q) \tag{35}$$

$v_A$ and $v_E$ are the L2d value in the American and European models; $v_E$ is the function that was referred to simply as $v$ in the previous section.

As described above, $v_E(-h,q)$ is calculated by projecting the terminal value back in time through a transformation, convolution, and back-transformation. Call this operation $\mathcal{P}_E(-h)$:

$$v_E(-h,q) = \mathcal{P}_E(-h) \cdot v_E(0,q) \tag{36}$$

If the time step $h$ is small, $v_A(-h,q)$ will be the same as $v_E$ for small values of $q$, i.e. poor environments, since the option to become a dauer is valuable in a poor environment, and it is unlikely that the environment will change from poor to good in the short remaining time $h$. Similarly, for large values of $q$ it is almost certain that the dauer option will not be exercised. Thus, for large $q$ the L2d has the value of an L2 of age $-h$, $V_{L3}qe^{-\lambda h}$. For intermediate quality environments, the American L2d chooses the option that maximizes its value:

$$v_A(-h,q) = \max(v_E(-h,q), V_{L3}qe^{-\lambda h})$$
$$= \max(\mathcal{P}_E(-h) \cdot v_E(0,q), V_{L3}qe^{-\lambda h}) \tag{37}$$
$$= \mathcal{P}_A(-h) \cdot v_A(0,q)$$

where $\mathcal{P}_A(-h)$ is defined by:

$$\mathcal{P}_A(-h) \cdot v_A(a,q) \equiv v_A(a-h,q)$$
$$= \max\left(\mathcal{P}_E(-h) \cdot v_A(a,q), V_{L3}qe^{\lambda(a-h)}\right) \tag{38}$$

This is an approximation, valid only if $h$ is so small that switching to the L2 pathway at some time between $-h$ and 0 will not be much better than switching at the better of $-h$ and 0. Of course, $v_A(-h,q)$ can be regarded as the terminal value of $v_A$ for an $h$ step back to $v_A(-2h,q)$. By iterating (38), $v_A(a,q)$ can be approximated all the way back to the L1 molt. In practice this calculation can't be carried out exactly as described for numerical stability reasons. However, a mathematically equivalent process in which the transformed function $y_A(a,u)$ is corrected at each step for the early exercise value can be made to work.

**Hybrid model.** In the American model, an L2d may discard the dauer option at any time; if it does so immediately after the L1 molt, it will incur no delay. However, Golden and Riddle [2] found that the real behavior is intermediate between American and European models: if an L2d was switched to favorable conditions immediately after the L1 molt, its development was delayed by only 2–3 hours. This result suggests that an L2d can change its mind before the molt, committing to reproductive development and growing at the faster reproductive rate. The mechanism of this delayed switch has recently been worked out [12].

I used a hybrid American/European model to evaluate the L2d for this case. In this model a worm that chooses the L2d pathway at the L1 molt must develop as an L2d at least up to a developmental age $a_{EE}$, the early exercise stage. At any age after $a_{EE}$ the worm may switch to the L2 pathway. In this model an L2d of age $a_{EE}$ or greater has exactly the same prospects as in the American model. For ages before $a_{EE}$, the hybrid model worm has a European option with expiration $a_{EE}$ on the American model worm at $a_{EE}$.

$$v_H(a,q) = \begin{cases} a \geq a_{EE} & v_A(a,q) \\ a < a_{EE} & \mathcal{P}_E(a-a_{EE}) \cdot v_A(a_{EE},q) \end{cases} \tag{39}$$

This is not the only way to model the minimum delay, nor is it obviously the correct one. For instance, it might seem more plausible that the worm can choose at any time to switch from L2d to L2, but that the switch can't be executed immediately; that some time is needed to unwind L2d development and restart L2 development. This model is very similar to the one described above: the worm develops as an L2d for some minimum time, then as an L2 after that. The difference is when it makes the choice. In the model described by (39) the worm makes its decision to switch later than in the slow switch model, where a worm must decide at the L1 molt in order to switch to the L2 pathway at $a_{EE}$. This early decision is a disadvantage, since it is based on less current information than in the hybrid model described. The hybrid model was used for the value calculations in the body of the paper, partly because it is computationally more tractable, but also because it gives a higher value for the L2d and is thus useful in establishing bounds.

## Future prospects

The model for growth and value described here has two obvious flaws. First, environment quality is modeled as a geometric Brownian motion (3). This means it can grow or diminish without bound, and in fact is expected to do one or the other in the long term. The real environment, however, is more stable. If things are very good, they will probably get worse; if things are very bad, they will probably get better. Second, growth and environment quality vary independently—they are governed by uncorrelated Brownian motions $W_1(t)$ and $W_2(t)$ (eqs (3) and (6)). In reality the worms will tend to grow faster in a good environment, slower in a poor environment. It is mathematically straightforward to modify the model so that the two Brownian motions are correlated, but this gives biologically absurd results, because it correlates the environment not with growth rate, but with worm age and condition $a(t)$.

These problems are related, and they can be fixed in the same way, by modeling the environment not as a geometric Brownian motion, but as the exponential of an Ornstein-Uhlenbeck process [27]. The Ornstein-Uhlenbeck process is mean-reverting and over long periods has a normal distribution. In the simplest version of such a model the growth rate of the worm (i.e., the rate of change of $a(t)$) is directly proportional to environment quality, which is an Ornstein-Uhlenbeck process. This model produces the product form (10) for value, not as an assumption, but as a derived consequence. $q(t)$ in this case depends on the difference between current environment quality and its long-term mean value. This model unfortunately is too simple, since it has no room for information about the environment with no immediate effect on growth (for instance, the smell of food, or pheromones that measure population density). It can be improved by modeling environment quality as the product of two factors, one that directly determines growth rate, and one that depends exponentially on a second Ornstein-Uhlenbeck process representing environmental information that doesn't immediately affect growth.

I have not fully developed these models and do not use them in this paper for two reasons. First, the PDEs they produce are more complicated, and I have not yet found an efficient way to solve them. Second and more important, these more complicated models have more free parameters, which I have no good way of estimating. For instance, the two time constants for mean-reversion of the two Ornstein-Uhlenbeck processes are important. Over times that are short compared to the mean-reversion time constant, an Ornstein-Uhlenbeck process looks like a Brownian motion. The Brownian motion models used here are reasonable approximations to the Ornstein-Uhlenbeck process models if mean-reversion times are longer than the duration of the L2d stage.

## Smart and Dumb worm strategies

The Smart and Dumb strategies are evaluated in a world where, over long times, volatility is zero 50% of the time and infinite 50% of the time. The Smart worm knows the current volatility and makes the optimal decision based on both volatility and environment quality. Its decisions and value can be understood based on Figure 2. At zero volatility, the L2d curve intersects the L1 curve at $q=0.3$, so the L1 chooses L2d if $q<0.3$ and L2 if $q>0.3$. Its value at the L1 molt is

$$v_{\text{smart},0}(q) = \max\left(V_{\text{d}}e^{\frac{\lambda a_{\text{L1molt}}}{\delta}}, V_{\text{L3}}qe^{\lambda a_{\text{L1molt}}}\right) \quad (40)$$

For $q<0.3$ this reduces to $V_{\text{d}}e^{\frac{\lambda a_{\text{L1molt}}}{\delta}}$, the discounted value of the dauer, and for $q>0.3$ it is $V_{\text{L3}}qe^{\lambda a_{\text{L1molt}}}$, the discounted value of the L3.

At infinite volatility, the intersection and hence the cutoff occurs at $q=2.4$. Its value is

$$v_{\text{smart},\infty}(q) =$$
$$\max\left(V_{\text{d}}e^{\frac{\lambda a_{\text{L1molt}}}{\delta}} + V_{\text{L3}}qe^{\lambda\left(\frac{a_{\text{L1molt}}-a_{\text{EE}}}{\delta}+a_{\text{EE}}\right)}, V_{\text{L3}}qe^{\lambda a_{\text{L1molt}}}\right) \quad (41)$$

In this expression the second term of the first argument is the value of a worm that chooses L2d at the L1 molt, then, at $a_{\text{EE}}$ switches to the L2 pathway.

Finally, the value of the smart worm is

$$v_{\text{smart}}(q) = 0.5 v_{\text{smart},0}(q) + 0.5 v_{\text{smart},\infty}(q) \quad (42)$$

This is piecewise linear with breakpoints at 0.3 and 2.4.

The Dumb worm doesn't know the current volatility, and must base its decision solely on $q$. If it chooses L2d and the volatility is 0, its value is given by (40); $v_{\text{dumbL2d},0}(q) = v_{\text{smart},0}(q)$. If the volatility is infinite,

$$v_{\text{dumbL2d},\infty}(q) = V_{\text{d}}e^{\frac{\lambda a_{\text{L1molt}}}{\delta}} + V_{\text{L3}}qe^{\lambda\left(\frac{a_{\text{L1molt}}-a_{\text{EE}}}{\delta}+a_{\text{EE}}\right)} \quad (43)$$

Thus, the long-term average value of the Dumb L2d is

$$v_{\text{dumbL2d}}(q) = 0.5 v_{\text{dumbL2d},0}(q) + 0.5 v_{\text{dumbL2d},\infty}(q) \quad (44)$$

This is piecewise linear with a single breakpoint at 0.34. The breakpoint is different from that for the Smart worm because the Dumb worm, not knowing the volatility, may choose the L2d pathway at zero volatility even when it is not optimal, i.e. at $q>0.3$. By the time it reaches $a_{\text{EE}}$ it will discover its mistake and, if $q>0.34$, will switch to the L2 pathway, cutting its losses. If, however, $q<0.34$, the optimum is now to become a dauer. This cutoff is different from the cutoff the smart worm makes at zero volatility, because the dumb worm has paid part of the price for becoming a dauer. Its remaining cost to become a dauer is now a 4 h delay rather than a 7 h delay, so the dauer option is more attractive.

The value of the L2 is by definition independent of volatility

$$v_{\text{dumbL2}}(q) = V_{\text{L3}}qe^{\lambda a_{\text{L1molt}}} \quad (45)$$

Finally, the value of the Dumb worm is

$$v_{\text{dumb}}(q) = \max(v_{\text{dumbL2d}}(q), v_{\text{dumbL2}}(q)) \quad (46)$$

The Dumb worm cutoff, the intersection of $v_{\text{dumbL2d}}(q)$ and $v_{\text{dumbL2}}(q)$, is at $q=1.2$. The difference between $v_{\text{smart}}(q)$ and $v_{\text{dumb}}(q)$, plotted in Figure 4B, thus has breakpoints at 0.3, 0.34, 1.2, and 2.4

$$v_{\text{smart}}(q) - v_{\text{dumb}}(q) = \begin{cases} q<0.3 & 0 \\ 0.3<q<0.34 & 0.42q-0.13 \\ 0.34<q<1.2 & 0.05q \\ 1.2<q<2.4 & 0.13-0.05q \\ 2.4<q & 0 \end{cases} \quad (47)$$

## Parameter estimation

While the qualitative conclusions of this analysis are robust to numerical assumptions, it is necessary to insert specific numerical values for the parameters to estimate quantitative consequences. These parameters are not fixed. Some are known to be subject to genetic control, and the same can be presumed of the rest. To estimate the consequences of uncertainty on value, I attempted to estimate likely values of these parameters in the current *C elegans* population.

**Data sources.** The most important inputs into the model are measured times of developmental events and rates. Different papers report different values, in part because they were measured at different temperatures, the most common being 20°C and 25°C. By matching the times of specific events such as hatching and the molts, I adjusted all times and rates to a common standard, choosing Hodgkin and Barnes [7] as the base 20°C life history to which others were matched.

Data come from the following sources:

- Hodgkin and Barnes [7]: Life cycle duration, wild-type and *tra-3* brood sizes and progeny production rates, sperm production rate.
- Cutter [28]: Alternative brood size and sperm production rate.
- Wood [29]: egg-laying, hatch, and molt times.
- Golden and Riddle [30]: minimum and maximum L2d durations, time to form and recover from dauer.
- Knight et al. [31], and data provided by Armand Leroi (personal communication): body volumes and growth curves.

To calculate $v(a,q)$, numerical values for parameters $\alpha$, $\lambda$, $\delta$, $\sigma$, $a_{EE}$, $V_d$, and $V_{L3}$ are needed.

**σ: volatility.** I did not attempt to estimate volatility, but left it as a free parameter in the simulations. The volatility $\sigma$ is the root-mean-square rate of change of the natural logarithm of environment quality with the square root of time and has units of $s^{-\frac{1}{2}}$. Since the meaning of numerical values in these units is difficult to interpret, I instead plot a derived quantity I call "uncertainty":

$$\text{uncertainty} = e^{\sigma\sqrt{T}} - 1 \qquad (48)$$

$T$ is the duration of the L2d stage (the modal duration in variable growth rate models), and $e^{\sigma\sqrt{T}}$ the factor by which environment quality changes in this time. 1 is subtracted so that a completely predictable environment has uncertainty 0. An uncertainty of 100 means that the environment at the end of the L2d is expected most often to be between 101 times better and 101 times worse than at the beginning.

**α: growth speed.** $\alpha$ is the drift in the Brownian motion growth model. This parameter sets the time scale. It determines how long a worm that successfully grows from age $a_1$ to age $a_2$ takes to do it. The values of $a$ at which particular events such as molts occur are defined to be the times after fertilization under ideal laboratory conditions at $20°C$, so ideal conditions would correspond to $\alpha = 1$. Growth rate depends on conditions, especially food availability and quality [32]. Thus, one would expect the growth rate in the wild to be lower. On the other hand, food deprivation more severe than that which slows growth 3–4-fold leads to larval death (unpublished observations). $\alpha$ should thus be chosen so that most worms take between a and 4a hours to advance in age by $a$. In the simulations presented here I choose $\alpha$ so that the typical (modal) worm requires 32.8 h for L2d development, twice the time required under ideal conditions.

**δ: L2d delay.** The L2d matures more slowly than the L2 [30], which results in delayed development. This is accounted for by the factor $\delta$ in the SDE for L2d growth. It is the L2 duration divided by the maximum L2d duration, $\delta = (8.8h)/(16.4h) = 0.54$.

**$a_{EE}$: L2d early exercise time.** Golden and Riddle [30] report that a worm grown under dauer-inducing conditions until the L1 molt, then shifted to conditions that favor reproductive development was delayed in development by 2–3 hours, the average of which corresponds to 3.2 h after correction for temperature. If a worm exercises the option to switch to the L2 pathway at developmental age $a$, it will develop at rate $\delta$ from the L1 molt until it reaches a, then at rate 1 from a to the L2 molt. The total time required is thus:

$$t = \frac{a - a_{L1molt}}{\delta} + \frac{a_{L2molt} - a}{1} \qquad (49)$$

Substituting $a_{L2molt} = 0$, $a_{L1molt} = -8.8$, $\delta = 0.54$, $t = 8.8 + 3.2 = 12.0$ and solving for $a$ shows that a worm that experienced a 3.2 h delay must have switched from L2d to L2 pathway at $a = -5.1h$; i.e., it spent 7 h as an L2d, during which it advanced as far in development as an L2 would in 3.8 h, followed by 5 h as an L2. Since this was the minimum developmental delay, $a_{EE} = -5.1h$ is the earliest time at which an L2d may exercise the option to switch to L2.

**λ: discount rate.** The discount rate, a key parameter, measures how rapidly value increases with $a$, $v \propto e^{\lambda a}$. I estimated this in three ways.

*Method 1: larval growth*: The main business of a hermaphrodite larva is to eat and grow in size, in order to eventually become an adult with a large intestine and gonad that can support the manufacture of eggs from food. It is a reasonable guess, then, that the value of a larva is proportional to its size. Larval growth is approximately exponential [31]. Using size measurements helpfully provided by Armand Leroi (personal communication), I estimated the rate of larval growth to be $\lambda_g = 0.064h^{-1}$, corresponding to a doubling time of 10.9 h.

*Method 2: reproductive rate*: In the long run, the value of any worm lies in its future progeny. The total value of future progeny, discounted to the present, should equal the value of the worm.

$$v = \int_0^\infty p(a)e^{-\lambda_r a}da \qquad (50)$$

Here $p(a)$ is the rate of production of eggs by a worm of age $a$ and $\lambda_r$ is the reproductive rate estimate of the discount rate. I neglect mortality, since it is very low before or during the reproductive period under laboratory conditions. If age 0 is fertilization and the newly fertilized egg is the accounting unit, $v = 1$ and (50) becomes the Euler-Lotka equation.

Hodgkin and Barnes [7] reported that an egg takes 64.4 h to mature to the point at which it produces its first egg. Thereafter it produces 5.3 per hour until the sperm supply runs out at 327. (50) thus becomes

$$1 = \int_{64.4}^{64.4 + 327/5.3} 5.3e^{-\lambda_r a}da \qquad (51)$$

This is solved by $\lambda_r = 0.068h^{-1}$, doubling time 10.3 h. I have observed that a single worm placed on a standard bacteria-seeded 6 cm plate exhausts the food and gives rise to between 100,000 and 200,000 descendants in 7 days (unpublished data), consistent with a reproductive rate in this range. Chen et al. [33] and Muschiol et al. [34] have published estimates of reproductive rate in the same ballpark, 0.042 h⁻¹ and 0.057 h⁻¹, respectively. Unfortunately, I couldn't use these estimates directly, since they sample at rather long intervals (24 h for Chen et al. [33], 6 h for Muschiol et al. [34]) and do not report the timing of developmental milestones that can be used to standardize age.

*Method 3: sperm production*: The two previous methods both estimated the rate of increase of value under laboratory conditions where food is not limiting. In the wild it is likely that worms frequently have limited resources and thus grow more slowly. This is accounted for in the model: a is assumed to increase at a slower rate in nature than in the lab (see "$_\alpha$: growth speed" estimate above), and future value is discounted by a rather than by time. However, it is also true that when food is limited worms are smaller and produce fewer progeny [35]. Thus, the discount rate in nature may be lower than the previous two lab-based estimates.

Hodgkin and Barnes [7] showed that a *tra-3* mutant that increases sperm production and thus produces 499 progeny delays progeny production by 2.6 hours. Despite the larger brood, this mutant has a slower population growth rate than wild type. In fact, these numbers predict that *tra-3* worms should grow 3.01% more slowly than wild-type, in agreement with the measured difference of 2.81±0.62% in eating races, suggesting that Hodgkin and Barnes correctly identified the reason for slower growth.

Surprisingly, however, the schedule of wild-type sperm production is not optimal for a discount rate of 0.068 h$^{-1}$. Comparison of wild-type and *tra-3* suggests that the production of each sperm delays egg production $2.6/(499-327) = 0.015$ h. This places the start of sperm production at $64.4-0.015 \times 327 = 59.5$ h. The value of a worm that makes $n$ sperm is thus estimated to be

$$v(n,\lambda) = \int_{59.5+0.015n}^{59.5+0.015n+n/5.3} 5.3 e^{-\lambda a} da$$
$$= 5.3 e^{-59.5\lambda} \frac{e^{-0.015n\lambda} - e^{-0.20n\lambda}}{\lambda}$$
(52)

(This equation is valid only for sperm production that delays egg production. As Cutter [28] has pointed out, many sperm are produced before adulthood and their production may have no effect on the timing of oogenesis. This consideration has no effect on the calculations that follow.) With $\lambda = 0.068 h^{-1}$, $v(n,\lambda)$ is maximized not at 327, but 204 sperm, at a maximum value of 1.07 eggs. One possible explanation is that the discount rate in nature is lower than 0.068 h$^{-1}$. By assuming that sperm production is approximately optimal for the discount rate in nature, I can estimate that rate. The optimal $n$ is that for which $\frac{\partial}{\partial n}v(n,\lambda_s) = e^{-59.5\lambda_s}(1.08e^{-0.20n\lambda_s} - 0.080e^{-0.015n\lambda_s}) = 0$. Letting $n = 327$ and solving for $\lambda_s$ gives $\lambda_s = 0.042 h^{-1}$, doubling time 16.4 h. Using Cutter's measured sperm production rate of 23.6 h$^{-1}$, the discount rate estimated by this method is instead 0.027 h$^{-1}$, doubling time 36.4 h. Here I prefer the rate estimated from Hodgkin and Barnes's data, as it's based on progeny timing, the most directly relevant numbers. However, there is no obvious reason why these two numbers should differ, and I suggest the discrepancy between 0.027 and 0.042 is a measure of the likely error in this estimate.

*Summary*: Three distinct methods give estimates of the discount rate $\lambda$ ranging from 0.027 h$^{-1}$ to 0.068 h$^{-1}$. The method I consider best gives $\lambda = 0.042 h^{-1}$, corresponding to a doubling time of 16.4 h.

**$V_d$, $V_{L3}$: new dauer and L3 values.** $V_d$ and $V_{L3}$ are expressed in terms of $\lambda$. I use a mature dauer as my unit of account. A dauer needs 14.5 h to recover to an L4 [2, corrected for temperature], so the value of an L4 right after the dauer molt is $V_{L4} = e^{14.5\lambda}$. Assuming that the L4 resulting from the dauer molt and the L4 resulting from the L3 molt are equivalent, $V_{L3}$ can be estimated by discounting the L4 value by the length of the L3 stage, 9.5 h, to give $V_{L3} = e^{(14.5-9.5)\lambda} = e^{5.1\lambda}$.

$V_d$ is the value of a newly formed dauer, right after the L2d molt. This is not 1, because full maturation of the dauer takes 15.9 h and is irreversible, once begun [2]. Thus $V_d = e^{-15.9\lambda}$.

## Software and calculations

Calculations were done in *Mathematica* (Wolfram Research, Inc). All calculations are included in Dataset S1 as *Mathematica* notebooks.

## Supporting Information

**Dataset S1 Calculations.** Calculations were done in Wolfram *Mathematica*. This dataset contains *Mathematica* notebooks that carry out the calculations and explain them in detail. There are two principle notebooks, amer_model_v6.nb and calculations_v2.nb. Files init_v3.m, init_v3.nb, itoCalculus_v6.m, itoCalculus_v6.nb, and Shreve.m contain supporting code necessary for amer_model_v6.nb and calculations_v2.nb to evaluate. PDF printouts of the two main notebooks are also provided so that they can be read without *Mathematica*.
(ZIP)

## Acknowledgments

Michael van Breda and Thomas Thibodeau first introduced me to quantitative finance. Armand Leroi provided unpublished larval growth data. Youngjai You, Alex Artyukhin, and Pauline Fotopoulos commented on the manuscript.

## Author Contributions

Conceived and designed the experiments: LA. Performed the experiments: LA. Analyzed the data: LA. Contributed to the writing of the manuscript: LA.